\documentclass[aps,prl,twocolumn,amsfonts,amssymb,amsmath,floats,floatfix,showpacs,preprintnumbers,superscriptaddress]{revtex4-1}

\usepackage{graphicx}
\usepackage{color}
\usepackage{dcolumn}
\usepackage{bm}
\usepackage{hyperref}
\usepackage{relsize}

\hypersetup{
colorlinks=true,
urlcolor= blue,
citecolor=blue,
linkcolor= blue,
bookmarks=true,
bookmarksopen=false,
}

\usepackage{multirow}
\usepackage{amsmath}
\usepackage{braket}

\newcommand{\SR}{Sr$_2$RuO$_4$}
\newcommand{\dg}{$^{\circ}$}

\newcommand{\op}[1]{\ensuremath{\vec{#1}}}
\newcommand{\ex}[1]{\left\langle#1\right\rangle}
\renewcommand{\ex}[1]{\langle\,#1\,\rangle}

\newcommand{\parstate}{\Ket{d_{-1_z}^{\downarrow_z}\!,d_{+1_z}^{\uparrow_z}\!}}
\newcommand{\aparstate}{\Ket{d_{-1_z}^{\uparrow_z}\!,d_{+1_z}^{\downarrow_z}\!}}

\begin{document}

\title{Breakdown of singlets and triplets in \SR{} revealed by spin-resolved ARPES}

\author{C.\,N. Veenstra}
\author{Z.\,-H.\,Zhu}
\author{M.\,Raichle}
\affiliation{Department of Physics {\rm {\&}} Astronomy, University of British Columbia, Vancouver, British Columbia V6T\,1Z1, Canada}
\author{B.\,M.\,Ludbrook}
\affiliation{Department of Physics {\rm {\&}} Astronomy, University of British Columbia, Vancouver, British Columbia V6T\,1Z1, Canada}
\author{A.\,Nicolaou}
\affiliation{Department of Physics {\rm {\&}} Astronomy, University of British Columbia, Vancouver, British Columbia V6T\,1Z1, Canada}
\affiliation{Quantum Matter Institute, University of British Columbia, Vancouver, British Columbia V6T\,1Z4, Canada}
\author{B.\,Slomski}
\author{G.\,Landolt}
\affiliation{Physik-Institut, Winterthurerstrasse 190, Universitat Z\"urich-Irchel, CH-8057 Z\"urich, Switzerland}
\affiliation{Swiss Light Source, Paul Scherrer Institut, CH-5232 Villigen PSI, Switzerland}
\author{S.\,Kittaka}
\affiliation{Department of Physics, Graduate School of Science, Kyoto University, Kyoto 606-8502, Japan}
\affiliation{Institute for Solid State Physics, University of Tokyo, Kashiwa, Chiba 277-8581, Japan}
\author{Y.\,Maeno}
\affiliation{Department of Physics, Graduate School of Science, Kyoto University, Kyoto 606-8502, Japan}
\author{J.\,H.\,Dil}
\affiliation{Physik-Institut, Winterthurerstrasse 190, Universitat Z\"urich-Irchel, CH-8057 Z\"urich, Switzerland}
\affiliation{Swiss Light Source, Paul Scherrer Institut, CH-5232 Villigen PSI, Switzerland}
\author{I.\,S.\,Elfimov}
\affiliation{Department of Physics {\rm {\&}} Astronomy, University of British Columbia, Vancouver, British Columbia V6T\,1Z1, Canada}
\affiliation{Quantum Matter Institute, University of British Columbia, Vancouver, British Columbia V6T\,1Z4, Canada}
\author{M.\,W.\,Haverkort}
\affiliation{Department of Physics {\rm {\&}} Astronomy, University of British Columbia, Vancouver, British Columbia V6T\,1Z1, Canada}
\affiliation{Quantum Matter Institute, University of British Columbia, Vancouver, British Columbia V6T\,1Z4, Canada}
\affiliation{Max Planck Institute for Solid State Research, Heisenbergstra\ss{}e 1, 70569 Stuttgart, Germany}
\author{A.\,\,Damascelli}
\email{damascelli@physics.ubc.ca}
\affiliation{Department of Physics {\rm {\&}} Astronomy, University of British Columbia, Vancouver, British Columbia V6T\,1Z1, Canada}
\affiliation{Quantum Matter Institute, University of British Columbia, Vancouver, British Columbia V6T\,1Z4, Canada}

\date{\today}

\begin{abstract} 
Spin--orbit coupling has been conjectured to play a key role in the low-energy electronic structure of \SR{}. Using circularly polarized light combined with spin- and angle-resolved photoemission spectroscopy, we directly measure the value of the effective spin--orbit coupling to be 130$\pm$30\,meV. This is even larger than theoretically predicted and comparable to the energy splitting of the $d_{xy}$ and $d_{xz,yz}$ orbitals around the Fermi surface, resulting in a strongly momentum-dependent entanglement of spin and orbital character. As demonstrated by the spin expectation value $\ex{\vec{s_k} \cdot \vec{s_{-k}}}$ obtained for a pair of electrons with zero total momentum, the classification of the Cooper pairs in terms of pure singlets or triplets fundamentally breaks down, necessitating a description of the unconventional superconducting state of \SR{} in terms of these newly found spin-orbital entangled eigenstates. 
\end{abstract}


\maketitle

After a flurry of experimental activity \cite{Nature.394.558,Nature.396.242,Nature.396.658,Science_phase,PRL_Kerr}, \SR{} has become a hallmark candidate for spin-triplet chiral $p$-wave superconductivity, the electronic analogue of superfluid $^3$He \cite{sigrist,Nature.396.627,rmp}. 
\begin{figure*}[t!]
\includegraphics[width=1\linewidth]{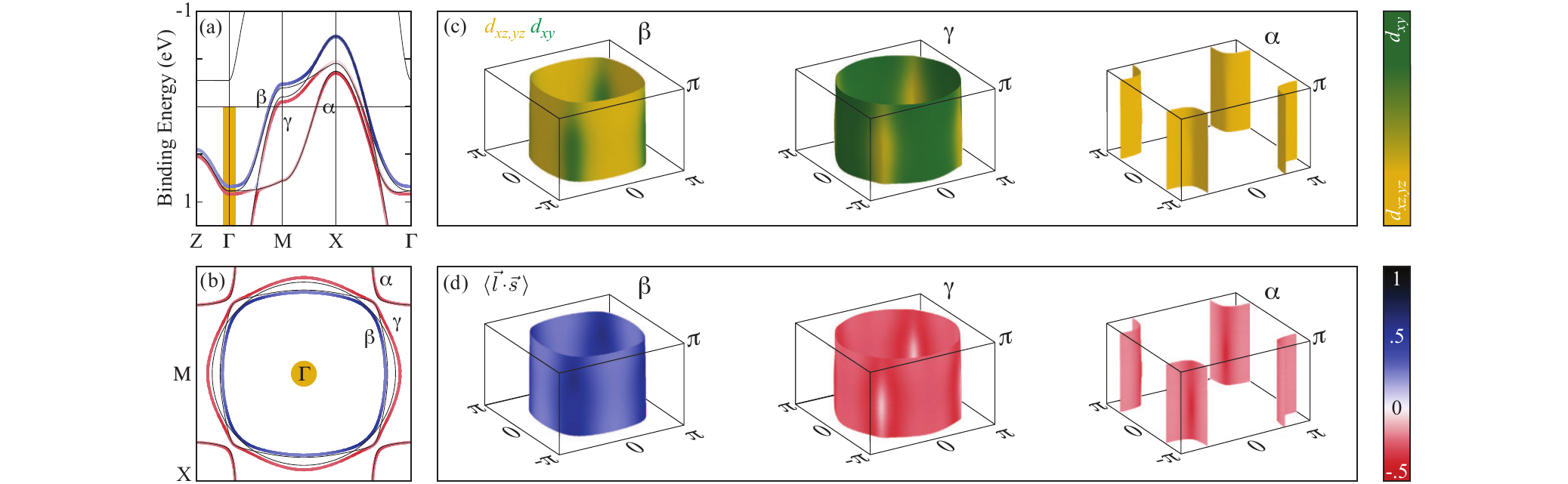}
\caption{(color online). (a) Band structure along the high-symmetry directions and (b) $k_z=0$ Fermi surface calculated without (thin black) and with (thick, color-coded to show $\ex{\op{l} \cdot \op{s}}$) the inclusion of SO coupling; at the $\Gamma$ point, the latter gives rise to a $\zeta_\textit{eff} \sim 90$\,meV splitting [note that Z$\equiv\!(0,0,\pi/c)$, $\Gamma\!\equiv\!(0,0,0)$, M$\equiv\!(\pi/a,0,0)$, X$\equiv\!(\pi/a,\pi/a,0)$]. (c,d) 3D Fermi surface sheets color-coded  to show (c) orbital character and (d) the expectation value $\ex{\op{l} \cdot \op{s}}$, in the first Brillouin zone \cite{extended}. The energy and momentum location of the spin-resolved ARPES spectra shown in Fig.\,\ref{SARPES} is marked in yellow in panels (a) and (b).}\label{BandStructure}
\end{figure*}
However, despite the apparent existence of such a pairing, some later experiments \cite{PhysRevLett.84.991,PhysRevB.80.174514,PhysRevLett.110.077003} do not fully support this conclusion, as they cannot be explained within a theoretical model using spin-triplet superconductivity alone \cite{JPSJ.81.011009}. A resolution might come from the inclusion of spin--orbit (SO) coupling, which has been conjectured to play a key role in the normal-state electronic structure \cite{PhysRevLett.101.026406} and may be important when describing superconductivity as well: by mixing the canonical spin eigenstates, the relativistic SO interaction might play a fundamental role beyond simply lifting the degeneracy of competing pairing states \cite{EPL_SO_Sigrist,raghu,PhysRevLett.101.026406,PhysRevLett.107.277003,arXiv1101.4656}. 

Thus far, the experimental study of SO coupling's effects on the electronic structure of \SR{} has been limited to the comparison of band calculations against angle-resolved photoemission spectroscopy (ARPES) \cite{PhysRevLett.101.026406,PhysRevLett.85.5194,PhysRevB.64.180502,MineNew,PhysRevLett.105.226406} -- no success has been obtained in observing experimentally either the strength of SO coupling or its implications for the mixing between spin and orbital descriptions. Here we probe this directly by performing spin-resolved ARPES \cite{LectNotePhys.697.95}, with circularly-polarized light: by using the angular momentum inherent in each photon -- along with electric-dipole selection rules \cite{Damascelli:physica} -- to generate spin-polarized photoemission from the SO mixed states. Combined with a novel spin- and orbitally-resolved ab-initio-based tight-binding (TB) modelling of the electronic structure \cite{supplementary}, these results demonstrate the presence of a non-trivial {\it spin--orbital entanglement} over much of the Fermi surface -- i.e. with no simple way of factoring the band states into the spatial and spin sectors. Most important, the analysis of the corresponding Cooper pair spin-eigenstates establishes the need for a description of superconductivity beyond the pure spin-triplet pairing.

In \SR{} the calculated effective SO coupling is small ($\zeta_\textit{eff} \sim 90$\,meV at the $\Gamma$ point) with respect to the bandwidth ($\sim$\,3\,eV) of the Ru-$t_{2g}$ orbitals, which define the $\alpha$, $\beta$, and $\gamma$ conduction bands. Nevertheless its influence always becomes important whenever bands would be degenerate in the absence of SO, either by symmetry or accidentally. This happens at several places in the three-dimensional (3D) Brillouin zone, as demonstrated in Fig.\,\ref{BandStructure}(a,b) where we show a comparison of the ab initio-TB band structure and Fermi surface calculated both with (color) and without (black) SO coupling included \cite{supplementary}. In the absence of SO, by symmetry the $d_{xz}$ and $d_{yz}$ bands would be degenerate along the entire $k_z$ momentum path from $\Gamma$ to Z [Fig.\,\ref{BandStructure}(a)]. Additionally, there are accidental degeneracies along the $k_z=0$ path from $\Gamma$ to X, where the bands of $d_{xz,yz}$ and $d_{xy}$ character all cross at momenta near ($2\pi/3,2\pi/3$) -- the exact location of which varies with $k_z$ but often occurs at the Fermi level [Fig.\,\ref{BandStructure}(a,b)]. At all these locations SO coupling naturally leads to a splitting [Fig.\,\ref{BandStructure}(a,b)] and mixing of the orbital character [Fig.\,\ref{BandStructure}(c)] for all three bands.

Interestingly, the effects of SO coupling are not limited to the regions around the non-relativistic degeneracies as, despite the large bandwidth, the Ru-$t_{2g}$ bands are often separated by energies small compared to the SO interaction. The predicted importance of the SO interaction can be directly visualized via the expectation value of $\vec{l}\cdot\vec{s}$ from our ab initio-TB modelling, with $\vec{l}$ and $\vec{s}$ being the orbital and spin angular momentum operators. A non-zero value of $\ex{\vec{l}\cdot\vec{s}}$ indicates complex orbital eigenstates that can be entangled with the spin. In this case, the wavefunction cannot be factorized into independent spin and orbital parts, as would be possible for a fully quenched angular momentum (for which $\ex{\vec{l}\cdot\vec{s}}=0$). The calculated $\ex{\vec{l}\cdot\vec{s}}$ is shown in Fig.\,\ref{BandStructure} for the high-symmetry dispersion (a), $k_z\!=\!0$ Fermi surface (b), and around the 3D Fermi sheets (d). This suggests SO coupling is important in \SR{} on almost the entire three-sheet Fermi surface \cite{note_ls}.

In order to probe the resulting internal spin--orbital structure of the electronic wavefunction, we turn to spin-resolved ARPES with circularly polarized light: with this technique the circular polarization of the light couples to the angular momentum of the states measured at a given $k$ point, while the spin is resolved directly. A similar approach, albeit without the angular and energy resolution needed to resolve the dispersive states belonging to the conduction band, has been used previously to generate spin-polarized photoemission from materials without a net magnetization, such as GaAs \cite{PhysRevB.13.5484} and Ca$_2$RuO$_4$ \cite{PhysRevLett.87.077202}. Here, by exploiting the electron-dipole selection rules for photoemission from conduction-band states selected via spin-resolved ARPES, we directly probe the internal SO structure of the normal state wavefunction (note that this study is done at $\sim\! 40$\,K, thus well above $T_c\!\simeq\!1.5$\,K).

To apply this technique on \SR{} we study the SO splitting at the $\Gamma$ point, $\vec{k}=(0,0,0)$, as highlighted in Fig.\,\ref{BandStructure}(a,b). This choice is dictated by the need of avoiding any intensity contamination from the well-known surface reconstruction of \SR{} \cite{PhysRevLett.85.5194,PhysRevB.64.180502,MineNew}, which leads to the detection of folded bands -- preventing a clean spin-ARPES study -- anywhere in the Brillouin zone except at the $\Gamma$ point \cite{supplementary}. In addition, as explained below, this choice selects the experimental geometry and initial-state wavefunctions that are the most straightforward to analyze, facilitating the direct measurement of both the SO interaction strength and the complex nature of the wavefunction. At this $k$-point, non-relativistic band structure calculations predict two degenerate bands of $d_{xz}$ and $d_{yz}$ character, with the $d_{xy}$ band far enough away that it can be ignored (i.e., at about 1.8-2.3\,eV higher binding energy, depending on the $k_z$ value). Here SO breaks the degeneracy by hybridizing these bands to form two states with a splitting of $\zeta_\textit{eff} \sim 90$\,meV: a lower binding-energy state with $z$-components of orbital and spin angular momentum {\it parallel} $\parstate$, and a higher binding-energy state where they are {\it antiparallel} $\aparstate$. Here ${\uparrow_z}$ represents spin, $d_{+1_z}\!\equiv\!\sqrt{1/2}(-d_{xz}\!-\!i d_{yz})$  has $m_{l_z}\!=\!1$, while $d_{-1_z}\!\equiv\!\sqrt{1/2}(d_{xz}\!-\!i d_{yz})$ has $m_{l_z}\!=\!-1$. Optical selection rules for the initial-to-final-state excitation with circularly polarized light dictate that both $\Delta \ell\!=\!\pm 1$ and  $\Delta m_{l}\!=\!\pm 1$. For $d$ orbitals the change in $\ell$ will favor the $d\!\rightarrow\!p$ over $d\!\rightarrow\!f$ transitions, owing to the cross-section at the photon energies used (24 and 56\,eV) and in particular the presence of a $d\!\rightarrow\!f$ Cooper minimum \cite{cooper} at 47\,eV for Ru$^{4+}$ (see also Supplemental Material \cite{supplementary}). The change in $m_l$ will depend on the circular polarization of the photon being right ($\oplus$) or left ($\ominus$). When a $\oplus$ ($\ominus$) photon is absorbed by the lower binding-energy parallel state $\parstate$, $m_{l_z}$ must increase (decrease) by one; but since an $|m_{l_z}|\!=\!2$ final state is forbidden in the favoured $p$ transition, electrons from the $d_{-1_z}^{\downarrow_z}$ ($d_{+1_z}^{\uparrow_z}$) half of the degenerate state will dominate, resulting in an effective $\downarrow_z$ ($\uparrow_z$) spin polarization. Similarly, photoemission from the higher binding-energy antiparallel state $\aparstate$ using $\oplus$ ($\ominus$) light will result in photoemission with the opposite spin polarization, $\uparrow_z$ ($\downarrow_z$). 

\begin{figure}[t!]
\includegraphics[width=1\linewidth]{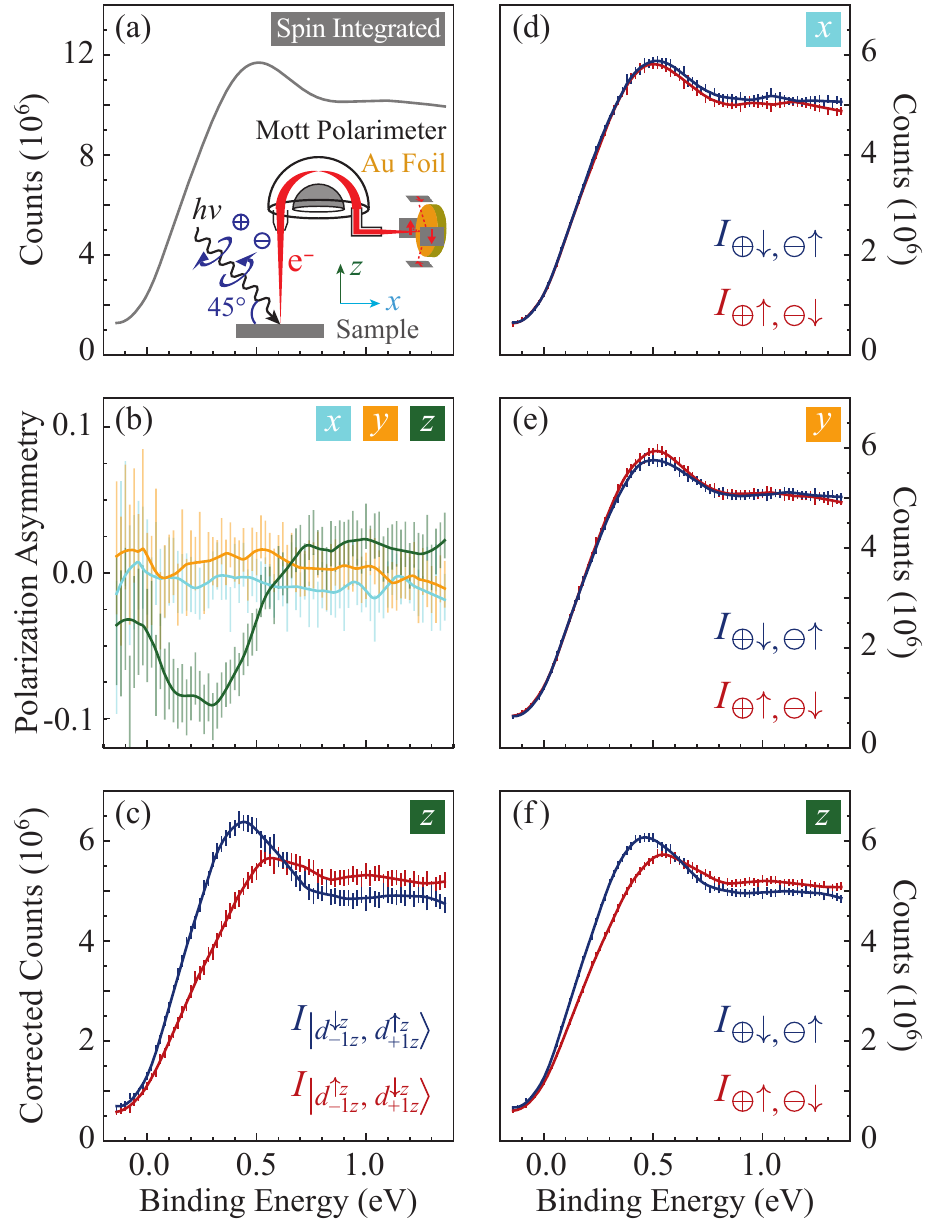}
\caption{(color online). (a) Spin-integrated ARPES data measured with 24\,eV photons at the $\Gamma$ point, as highlighted in Fig.\,\ref{BandStructure}. (b) Measured polarization asymmetry of the photoemitted electrons, and (d-f) corresponding spin-resolved ARPES intensities for $x$, $y$, and $z$ crystal axes, obtained with right ($\oplus$) or left ($\ominus$) circular polarization [see inset of (a) for experiment schematics]. (c) Intensity from each underlying state for the $z$ direction, corrected for light incident at 45\dg{} with respect to the spin--orbit quantization axis, as detailed in the Supplemental Material \cite{supplementary}. Vertical error bars represent statistical uncertainty based on number of counts in the Mott polarimeters, plotted at 95\% confidence, shown together with locally-weighted scatterplot smoothing fits \cite{supplementary}.
}\label{SARPES}
\end{figure}

In spin-integrated ARPES [Fig.\,\ref{SARPES}(a)], these $\Gamma$-point states are detected as a single broad feature with width $\sim\!400$\,meV \cite{supplementary}; however, it is possible to distinguish them by using circularly polarized light and observing the spin-polarization of the photoelectrons [see schematics in Fig.\,\ref{SARPES}(a)]. The experiment is repeated for both helicities of light, and the results combined to calculate the photoelectron {\it polarization asymmetry}, which eliminates possible experimental artefacts \cite{supplementary}. This polarization asymmetry is presented in Fig.\,\ref{SARPES}(b): it is zero along $x$ and $y$ crystal axes, and shows a clear wiggle as a function of energy along $z$, indicating that the photoelectrons have a photon-helicity-dependent spin-polarization only in the $z$-direction. By plotting the intensities corresponding to the observed photoelectron polarization asymmetry for each spatial dimension, Fig.\,\ref{SARPES}(d-f), we can directly resolve these states. For the $z$ direction in Fig.\,\ref{SARPES}(f) -- and in particular in Fig.\,\ref{SARPES}(c) where the data have been corrected for light incident at 45\dg{} with respect to the spin--orbit quantization axis \cite{supplementary} -- they become visible as two energy-split features: $\parstate$ photoemits $\downarrow_z$ ($\uparrow_z$) with $\oplus$ ($\ominus$) light, and is thus detected in $I_{\oplus\downarrow, \ominus\uparrow}$; similarly, $\aparstate$ is detected in $I_{\oplus \uparrow,\ominus\downarrow}$. Along the $x$ and $y$ directions in Fig.\,\ref{SARPES}(d,e), however, the spectra match the spin-integrated intensity in Fig.\,\ref{SARPES}(a) since the photoelectrons from both states have $\ex{s_x}\!=\!\ex{s_y}\!=\!0$ for both light helicities.
\begin{figure*}[t!]
\includegraphics[width=1\linewidth]{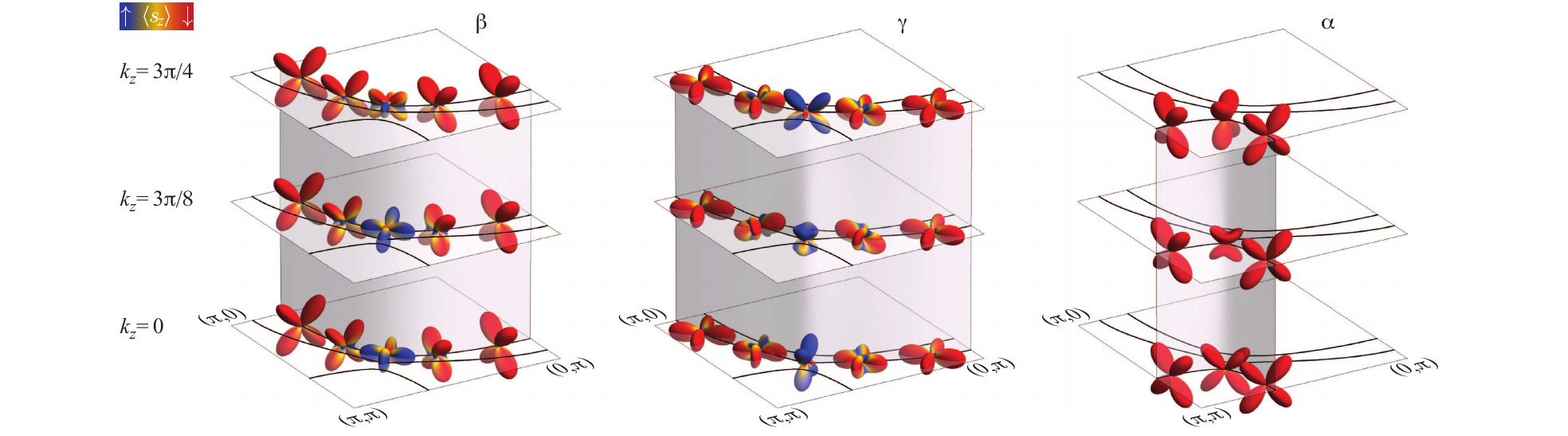}
\caption{(color online). Momentum-dependent Ru-$d$ orbital projection of the wavefunction for the $\beta$, $\gamma$, and $\alpha$ bands at selected momentum locations on the 3D Fermi surface. The surface color represents the momentum-dependent $s_z$ expectation value along the direction defined by the spherical $(\theta,\phi)$ angles, $\langle s_z \rangle_{(\theta,\phi)}$ \cite{supplementary}; as indicated by the color scale at upper left, blue/red correspond to spin $\uparrow$/$\downarrow$ for one state of the Kramers-degenerate pair (with the opposite spin state not shown \cite{kramer}). The strongly mixed colors on some of the orbital projection surfaces indicate strong, momentum-dependent spin--orbital entanglement.}\label{OrbitalSketch}
\end{figure*}
The splitting in the $z$ direction is observed with both 24\,eV and 56\,eV photons and its magnitude is $130\,\pm\,30$\,meV \cite{supplementary}, showing a possible enhancement compared to the predicted value $\zeta_\textit{eff}\!\sim\!90$\,meV.  Most  importantly, the existence of these two states, from which spin-polarized photoemission can be generated using circularly polarized light {\it in the $z$ direction only}, is clear experimental evidence of the importance of SO coupling in \SR{} and of its consequences for the complex nature of the normal-state wavefunctions. 

As discussed below, the most important of these consequences is the strong, momentum-dependent, spin--orbital entanglement of the eigenstates around the Fermi surface. This is illustrated in Fig.\,\ref{OrbitalSketch} by plotting the projection of the Bloch wavefunctions at the Fermi energy onto the Ru-$d$ orbital basis at different momenta \cite{supplementary}. The resulting projections are color-coded by the expectation value of the spin operator $\langle s_z \rangle_{(\theta,\phi)}$ for one half of the Kramers-degenerate pair (blue$=\!\uparrow$, red$=\!\downarrow$) \cite{kramer}. Along the edges of the Brillouin zone (M$-$X) where the bands are well-separated, we find particularly in the $\alpha$ band (far right panel in Fig.\,\ref{OrbitalSketch}) that the orbitals do not show strong entanglement: each orbital projection is associated with a single expectation value ({\it color}) of the spin operator. We also notice that in these areas the $\beta$ and $\alpha$ bands are of pure $d_{xz,yz}$ orbital character, and the $\gamma$ band of $d_{xy}$ (Fig.\,\ref{OrbitalSketch}). At these locations the wavefunction could be well approximated by the usual description as a product of independent spatial and spin components:
\begin{equation}
\psi(\mathbf{k}, \sigma)=\varphi(\mathbf{k}) \;
\phi^{\mathrm{spin}}_{\sigma}\, ,
\label{SimpleCooper}
\end{equation}
where $\varphi(\mathbf{k})$ and $\phi^{\mathrm{spin}}_{\sigma}$ are the spin and orbital eigenstates, and $\sigma$ the spin index. However, close to the zone diagonal, e.g. near the intersections of the Fermi sheets with $\Gamma\!-$X, this is not the case. Here we find strong orbital mixing for all bands and, especially in the $\gamma$ and $\beta$ bands, also strong entanglement between orbital and spin character: the orbitals are no longer associated with a uniform spin value; on the contrary, the latter can vary from fully up to fully down along a single orbital projection surface. Here the wavefunction cannot be written as in Eq.\,\ref{SimpleCooper}, and instead we must use the more general expression:
\begin{equation}
\psi(\mathbf{k}, \tilde{\sigma})=c_\uparrow \varphi_{\uparrow}(\mathbf{k})
\phi^{\mathrm{spin}}_{\uparrow} 
+
c_\downarrow \varphi_{\downarrow}(\mathbf{k})
\phi^{\mathrm{spin}}_{\downarrow} \, ,
\label{SOwavefunction}
\end{equation}
with $\tilde{\sigma}$ being the pseudo-spin index, and $c_{\uparrow,\downarrow}$ the prefactors of the momentum-dependent spin-orbital-entangled eigenstates. Eq.\,\ref{SOwavefunction} further illustrates the nature of the SO-induced entanglement: flipping the spin forces also a change of the orbital character. We note that, due to the nature of the band structure in \SR{}, this entanglement is strongly dependent on both $k_\parallel$ and $k_z$, despite the extremely weak $k_z$ dispersion along the Fermi surface.

A similar momentum and orbital dependence of the spin expectation value is responsible -- in topological insulators -- for the complex spin-texture of the Dirac fermions \cite{TI_RMP,jason,jasonnew}. In \SR{}, beyond the normal-state properties, it directly affects the description of superconductivity, as revealed by the inspection of the Cooper pair basic structure. Cooper assumed the two-particle wavefunction describing a superconducting electron pair to be of the form $\Psi(\mathbf{r}_1, \sigma_1, \mathbf{r}_2, \sigma_2)=\varphi(\mathbf{r}_1-\mathbf{r}_2) \phi^{\mathrm{spin}}_{\sigma_1,\sigma_2}$, with zero total momentum and the spin part being either singlet (total spin $S\!=\!0$) or triplet ($S\!=\!1$) \cite{PhysRev.108.1175}. This allows one to classify superconductors as a realization of singlet or triplet paired states. However, a fundamental assumption of this description is that one can write the wavefunction of each electron as a simple product of spatial and spin parts, which is not possible in the case of strong mixing between $\varphi_{\uparrow}(\mathbf{k})$ and $\varphi_{\downarrow}(\mathbf{k})$. Additionally, because of the strong 3D $k$-dependence of this entanglement in \SR{}, any transform to pseudo-spin would also necessarily be $k$-dependent -- negating the possibility of using the regular description under a pseudo-spin basis as might be done, e.g., for the heavy-fermion Ce compounds \cite{PhysRevLett.104.017002,heavy}.
\begin{figure}[b!]
\includegraphics[width=1\linewidth]{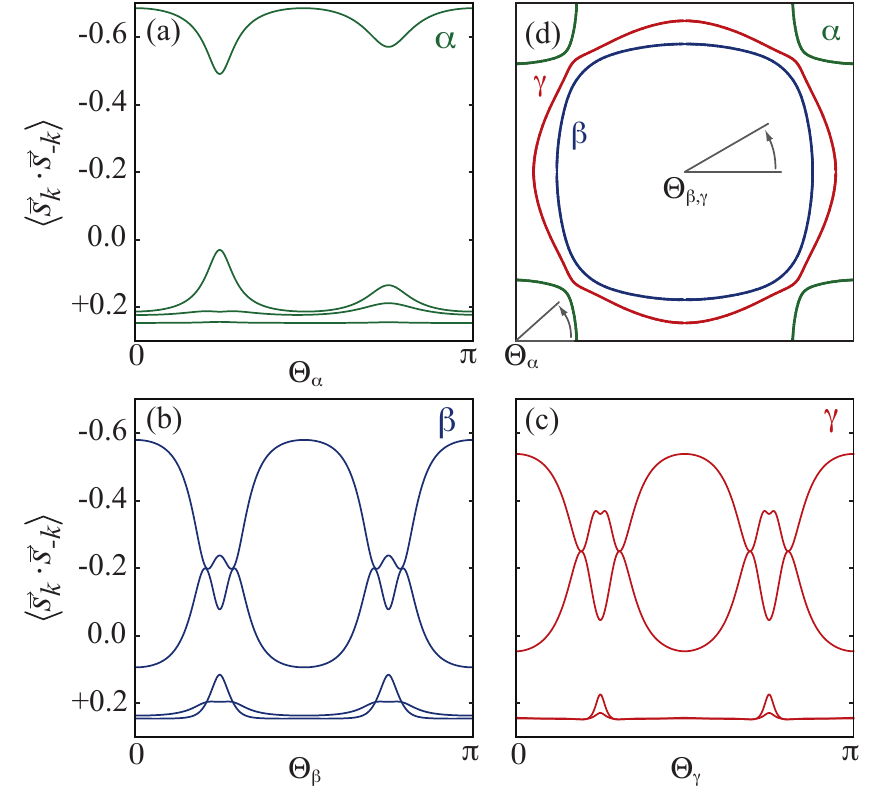}
\caption{(color online). Two-particle spin expectation value $\ex{\vec{s_k} \cdot \vec{s_{-k}}}$ for states with zero total momentum along the $k_z$=0 Fermi surface sheets for (a)  $\alpha$, (b) $\beta$, and (c) $\gamma$ bands. The $k_x$--$k_y$ plane location is defined by the angle $\Theta$ for each band, as illustrated in (d). The complete set of results for the full $k_z$ range is shown in Fig.\,5S  of the Supplemental Material \cite{supplementary}.}\label{sdotskz0}
\end{figure}
As a consequence, the classification of Cooper pairs in terms of singlets or triplets fundamentally breaks down for \SR{}. This is shown in Fig.\,\ref{sdotskz0} for $k_z$=0 (and in Fig.\,S5 of the Supplemental Material for the full $k_z$ range \cite{supplementary}), which presents the spin-eigenstates available to a pair of electrons with zero total momentum, as obtained from the expectation value $\ex{\vec{s_k} \cdot \vec{s_{-k}}}$, plotted versus the Fermi surface angle $\Theta$ defined in Fig.\,\ref{sdotskz0}(d). While familiar singlet and triplet states are seen off the zone diagonal for the $\alpha$ band (with $\ex{\vec{s_k} \cdot \vec{s_{-k}}}\!=\!-3/4$ and $1/4$, respectively), they are not available for either the $\beta$ or $\gamma$ bands, whose spin-eigensystem consists of a doublet and two singlets or -- depending on $\Theta$ -- two doublets.

Our findings mark a deviation from a pure spin-triplet pairing for \SR{}, since the only portion of the Fermi surface that might support it is contained within the $\alpha$ pocket, and suggest that superconductivity is yet more unconventional than has been assumed so far. This could explain a number of experimental observations at variance with a spin-triplet scenario, such as the extreme sensitivity to field angle of both the magnetic-field-induced second superconducting phase transition \cite{PhysRevLett.84.991} and also the suppression of the $ab$-plane upper critical field \cite{PhysRevB.80.174514}. These provide evidence for an additional magnetic anisotropy in the superconducting state, of which the entanglement of spin and orbit at the single-particle level would be the natural source. In this regard, it would be interesting to verify what of the chiral $p$-wave superconductor phenomenology \cite{Nature.394.558,Nature.396.242,Nature.396.658,Science_phase,PRL_Kerr}, and apparent conflict in experimental evidence \cite{PhysRevLett.84.991,PhysRevB.80.174514,PhysRevLett.110.077003}, would remain when re-evaluated in terms of entangled single-particle eigenstates.

We acknowledge R.J. Green, A. Kapitulnik, W.A. MacFarlane, G.A. Sawatzky, P.C.E. Stamp, and L.H. Tjeng for discussions. This work was supported by the Max Planck - UBC Centre for Quantum Materials (A.N., M.W.H.), the Killam, Alfred P. Sloan, Alexander von Humboldt, and NSERC's Steacie Memorial Fellowship Programs (A.D.), the Canada Research Chairs Program (A.D.), NSERC, CFI, CIFAR Quantum Materials, MEXT KAKENHI (No. 22103002), and the Deutsche Forschungsgemeinschaft through Forschergruppe FOR 1346.

\bibliography{SARPES_Sr214_PRL_revised}

\begin{thebibliography}{36}%
\makeatletter
\providecommand \@ifxundefined [1]{%
 \@ifx{#1\undefined}
}%
\providecommand \@ifnum [1]{%
 \ifnum #1\expandafter \@firstoftwo
 \else \expandafter \@secondoftwo
 \fi
}%
\providecommand \@ifx [1]{%
 \ifx #1\expandafter \@firstoftwo
 \else \expandafter \@secondoftwo
 \fi
}%
\providecommand \natexlab [1]{#1}%
\providecommand \enquote  [1]{``#1''}%
\providecommand \bibnamefont  [1]{#1}%
\providecommand \bibfnamefont [1]{#1}%
\providecommand \citenamefont [1]{#1}%
\providecommand \href@noop [0]{\@secondoftwo}%
\providecommand \href [0]{\begingroup \@sanitize@url \@href}%
\providecommand \@href[1]{\@@startlink{#1}\@@href}%
\providecommand \@@href[1]{\endgroup#1\@@endlink}%
\providecommand \@sanitize@url [0]{\catcode `\\12\catcode `\$12\catcode
  `\&12\catcode `\#12\catcode `\^12\catcode `\_12\catcode `\%12\relax}%
\providecommand \@@startlink[1]{}%
\providecommand \@@endlink[0]{}%
\providecommand \url  [0]{\begingroup\@sanitize@url \@url }%
\providecommand \@url [1]{\endgroup\@href {#1}{\urlprefix }}%
\providecommand \urlprefix  [0]{URL }%
\providecommand \Eprint [0]{\href }%
\providecommand \doibase [0]{http://dx.doi.org/}%
\providecommand \selectlanguage [0]{\@gobble}%
\providecommand \bibinfo  [0]{\@secondoftwo}%
\providecommand \bibfield  [0]{\@secondoftwo}%
\providecommand \translation [1]{[#1]}%
\providecommand \BibitemOpen [0]{}%
\providecommand \bibitemStop [0]{}%
\providecommand \bibitemNoStop [0]{.\EOS\space}%
\providecommand \EOS [0]{\spacefactor3000\relax}%
\providecommand \BibitemShut  [1]{\csname bibitem#1\endcsname}%
\let\auto@bib@innerbib\@empty
\bibitem [{\citenamefont {Luke}\ \emph {et~al.}(1998)\citenamefont {Luke},
  \citenamefont {Fudamoto}, \citenamefont {Kojima}, \citenamefont {Larkin},
  \citenamefont {Merrin}, \citenamefont {Nachumi}, \citenamefont {Uemura},
  \citenamefont {Maeno}, \citenamefont {Mao}, \citenamefont {Mori},
  \citenamefont {Nakamura},\ and\ \citenamefont {Sigrist}}]{Nature.394.558}%
  \BibitemOpen
  \bibfield  {author} {\bibinfo {author} {\bibfnamefont {G.~M.}\ \bibnamefont
  {Luke}}, \bibinfo {author} {\bibfnamefont {Y.}~\bibnamefont {Fudamoto}},
  \bibinfo {author} {\bibfnamefont {K.~M.}\ \bibnamefont {Kojima}}, \bibinfo
  {author} {\bibfnamefont {M.~I.}\ \bibnamefont {Larkin}}, \bibinfo {author}
  {\bibfnamefont {J.}~\bibnamefont {Merrin}}, \bibinfo {author} {\bibfnamefont
  {B.}~\bibnamefont {Nachumi}}, \bibinfo {author} {\bibfnamefont {Y.~J.}\
  \bibnamefont {Uemura}}, \bibinfo {author} {\bibfnamefont {Y.}~\bibnamefont
  {Maeno}}, \bibinfo {author} {\bibfnamefont {Z.~Q.}\ \bibnamefont {Mao}},
  \bibinfo {author} {\bibfnamefont {Y.}~\bibnamefont {Mori}}, \bibinfo {author}
  {\bibfnamefont {H.}~\bibnamefont {Nakamura}}, \ and\ \bibinfo {author}
  {\bibfnamefont {M.}~\bibnamefont {Sigrist}},\ }\href {\doibase 10.1038/29038}
  {\bibfield  {journal} {\bibinfo  {journal} {Nature}\ }\textbf {\bibinfo
  {volume} {394}},\ \bibinfo {pages} {558} (\bibinfo {year}
  {1998})}\BibitemShut {NoStop}%
\bibitem [{\citenamefont {Riseman}\ \emph {et~al.}(1998)\citenamefont
  {Riseman}, \citenamefont {Kealey}, \citenamefont {Forgan}, \citenamefont
  {Mackenzie}, \citenamefont {Galvin}, \citenamefont {Tyler}, \citenamefont
  {Lee}, \citenamefont {Ager}, \citenamefont {Paul}, \citenamefont {Aegerter},
  \citenamefont {Cubitt}, \citenamefont {Mao}, \citenamefont {Akima},\ and\
  \citenamefont {Maeno}}]{Nature.396.242}%
  \BibitemOpen
  \bibfield  {author} {\bibinfo {author} {\bibfnamefont {T.~M.}\ \bibnamefont
  {Riseman}}, \bibinfo {author} {\bibfnamefont {P.~G.}\ \bibnamefont {Kealey}},
  \bibinfo {author} {\bibfnamefont {E.~M.}\ \bibnamefont {Forgan}}, \bibinfo
  {author} {\bibfnamefont {A.~P.}\ \bibnamefont {Mackenzie}}, \bibinfo {author}
  {\bibfnamefont {L.~M.}\ \bibnamefont {Galvin}}, \bibinfo {author}
  {\bibfnamefont {A.~W.}\ \bibnamefont {Tyler}}, \bibinfo {author}
  {\bibfnamefont {S.~L.}\ \bibnamefont {Lee}}, \bibinfo {author} {\bibfnamefont
  {C.}~\bibnamefont {Ager}}, \bibinfo {author} {\bibfnamefont {D.~M.}\
  \bibnamefont {Paul}}, \bibinfo {author} {\bibfnamefont {C.~M.}\ \bibnamefont
  {Aegerter}}, \bibinfo {author} {\bibfnamefont {R.}~\bibnamefont {Cubitt}},
  \bibinfo {author} {\bibfnamefont {Z.~Q.}\ \bibnamefont {Mao}}, \bibinfo
  {author} {\bibfnamefont {T.}~\bibnamefont {Akima}}, \ and\ \bibinfo {author}
  {\bibfnamefont {Y.}~\bibnamefont {Maeno}},\ }\href {\doibase 10.1038/24335}
  {\bibfield  {journal} {\bibinfo  {journal} {Nature}\ }\textbf {\bibinfo
  {volume} {396}},\ \bibinfo {pages} {242} (\bibinfo {year}
  {1998})}\BibitemShut {NoStop}%
\bibitem [{\citenamefont {Ishida}\ \emph {et~al.}(1998)\citenamefont {Ishida},
  \citenamefont {Mukuda}, \citenamefont {Kitaoka}, \citenamefont {Asayama},
  \citenamefont {Mao}, \citenamefont {Mori},\ and\ \citenamefont
  {Maeno}}]{Nature.396.658}%
  \BibitemOpen
  \bibfield  {author} {\bibinfo {author} {\bibfnamefont {K.}~\bibnamefont
  {Ishida}}, \bibinfo {author} {\bibfnamefont {H.}~\bibnamefont {Mukuda}},
  \bibinfo {author} {\bibfnamefont {Y.}~\bibnamefont {Kitaoka}}, \bibinfo
  {author} {\bibfnamefont {K.}~\bibnamefont {Asayama}}, \bibinfo {author}
  {\bibfnamefont {Z.~Q.}\ \bibnamefont {Mao}}, \bibinfo {author} {\bibfnamefont
  {Y.}~\bibnamefont {Mori}}, \ and\ \bibinfo {author} {\bibfnamefont
  {Y.}~\bibnamefont {Maeno}},\ }\href {\doibase 10.1038/25315} {\bibfield
  {journal} {\bibinfo  {journal} {Nature}\ }\textbf {\bibinfo {volume} {396}},\
  \bibinfo {pages} {658} (\bibinfo {year} {1998})}\BibitemShut {NoStop}%
\bibitem [{\citenamefont {Nelson}\ \emph {et~al.}(2004)\citenamefont {Nelson},
  \citenamefont {Mao}, \citenamefont {Maeno},\ and\ \citenamefont
  {Liu}}]{Science_phase}%
  \BibitemOpen
  \bibfield  {author} {\bibinfo {author} {\bibfnamefont {K.~D.}\ \bibnamefont
  {Nelson}}, \bibinfo {author} {\bibfnamefont {Z.~Q.}\ \bibnamefont {Mao}},
  \bibinfo {author} {\bibfnamefont {Y.}~\bibnamefont {Maeno}}, \ and\ \bibinfo
  {author} {\bibfnamefont {Y.}~\bibnamefont {Liu}},\ }\href {\doibase
  10.1126/science.1103881} {\bibfield  {journal} {\bibinfo  {journal}
  {Science}\ }\textbf {\bibinfo {volume} {306}},\ \bibinfo {pages} {1151}
  (\bibinfo {year} {2004})}\BibitemShut {NoStop}%
\bibitem [{\citenamefont {Xia}\ \emph {et~al.}(2006)\citenamefont {Xia},
  \citenamefont {Maeno}, \citenamefont {Beyersdorf}, \citenamefont {Fejer},\
  and\ \citenamefont {Kapitulnik}}]{PRL_Kerr}%
  \BibitemOpen
  \bibfield  {author} {\bibinfo {author} {\bibfnamefont {J.}~\bibnamefont
  {Xia}}, \bibinfo {author} {\bibfnamefont {Y.}~\bibnamefont {Maeno}}, \bibinfo
  {author} {\bibfnamefont {P.~T.}\ \bibnamefont {Beyersdorf}}, \bibinfo
  {author} {\bibfnamefont {M.~M.}\ \bibnamefont {Fejer}}, \ and\ \bibinfo
  {author} {\bibfnamefont {A.}~\bibnamefont {Kapitulnik}},\ }\href {\doibase
  10.1103/PhysRevLett.97.167002} {\bibfield  {journal} {\bibinfo  {journal}
  {Phys. Rev. Lett.}\ }\textbf {\bibinfo {volume} {97}},\ \bibinfo {pages}
  {167002} (\bibinfo {year} {2006})}\BibitemShut {NoStop}%
\bibitem [{\citenamefont {Rice}\ and\ \citenamefont {Sigrist}(1995)}]{sigrist}%
  \BibitemOpen
  \bibfield  {author} {\bibinfo {author} {\bibfnamefont {T.~M.}\ \bibnamefont
  {Rice}}\ and\ \bibinfo {author} {\bibfnamefont {M.}~\bibnamefont {Sigrist}},\
  }\href@noop {} {\bibfield  {journal} {\bibinfo  {journal} {J. Phys.: Condens.
  Matter}\ }\textbf {\bibinfo {volume} {7}},\ \bibinfo {pages} {L643} (\bibinfo
  {year} {1995})}\BibitemShut {NoStop}%
\bibitem [{\citenamefont {Rice}(1998)}]{Nature.396.627}%
  \BibitemOpen
  \bibfield  {author} {\bibinfo {author} {\bibfnamefont {M.}~\bibnamefont
  {Rice}},\ }\href {\doibase 10.1038/25241} {\bibfield  {journal} {\bibinfo
  {journal} {Nature}\ }\textbf {\bibinfo {volume} {396}},\ \bibinfo {pages}
  {627} (\bibinfo {year} {1998})}\BibitemShut {NoStop}%
\bibitem [{\citenamefont {Mackenzie}\ and\ \citenamefont {Maeno}(2003)}]{rmp}%
  \BibitemOpen
  \bibfield  {author} {\bibinfo {author} {\bibfnamefont {A.~P.}\ \bibnamefont
  {Mackenzie}}\ and\ \bibinfo {author} {\bibfnamefont {Y.}~\bibnamefont
  {Maeno}},\ }\href@noop {} {\bibfield  {journal} {\bibinfo  {journal} {Rev.
  Mod. Phys.}\ }\textbf {\bibinfo {volume} {75}},\ \bibinfo {pages} {657}
  (\bibinfo {year} {2003})}\BibitemShut {NoStop}%
\bibitem [{ext()}]{extended}%
  \BibitemOpen
  \href@noop {} {\bibinfo  {journal} {See also Fig.\,S1 in Supplemental
  Material, which presents the Fermi surface in the conventional Brillouin zone
  of the body-centered tetragonal unit cell of
  ${\mathrm{Sr}}_{2}{\mathrm{RuO}}_{4}$}\ }\BibitemShut {NoStop}%
\bibitem [{\citenamefont {Mao}\ \emph {et~al.}(2000)\citenamefont {Mao},
  \citenamefont {Maeno}, \citenamefont {NishiZaki}, \citenamefont {Akima},\
  and\ \citenamefont {Ishiguro}}]{PhysRevLett.84.991}%
  \BibitemOpen
\bibfield  {journal} {  }\bibfield  {author} {\bibinfo {author} {\bibfnamefont
  {Z.~Q.}\ \bibnamefont {Mao}}, \bibinfo {author} {\bibfnamefont
  {Y.}~\bibnamefont {Maeno}}, \bibinfo {author} {\bibfnamefont
  {S.}~\bibnamefont {NishiZaki}}, \bibinfo {author} {\bibfnamefont
  {T.}~\bibnamefont {Akima}}, \ and\ \bibinfo {author} {\bibfnamefont
  {T.}~\bibnamefont {Ishiguro}},\ }\href {\doibase 10.1103/PhysRevLett.84.991}
  {\bibfield  {journal} {\bibinfo  {journal} {Phys. Rev. Lett.}\ }\textbf
  {\bibinfo {volume} {84}},\ \bibinfo {pages} {991} (\bibinfo {year}
  {2000})}\BibitemShut {NoStop}%
\bibitem [{\citenamefont {Kittaka}\ \emph {et~al.}(2009)\citenamefont
  {Kittaka}, \citenamefont {Nakamura}, \citenamefont {Aono}, \citenamefont
  {Yonezawa}, \citenamefont {Ishida},\ and\ \citenamefont
  {Maeno}}]{PhysRevB.80.174514}%
  \BibitemOpen
  \bibfield  {author} {\bibinfo {author} {\bibfnamefont {S.}~\bibnamefont
  {Kittaka}}, \bibinfo {author} {\bibfnamefont {T.}~\bibnamefont {Nakamura}},
  \bibinfo {author} {\bibfnamefont {Y.}~\bibnamefont {Aono}}, \bibinfo {author}
  {\bibfnamefont {S.}~\bibnamefont {Yonezawa}}, \bibinfo {author}
  {\bibfnamefont {K.}~\bibnamefont {Ishida}}, \ and\ \bibinfo {author}
  {\bibfnamefont {Y.}~\bibnamefont {Maeno}},\ }\href {\doibase
  10.1103/PhysRevB.80.174514} {\bibfield  {journal} {\bibinfo  {journal} {Phys.
  Rev. B}\ }\textbf {\bibinfo {volume} {80}},\ \bibinfo {pages} {174514}
  (\bibinfo {year} {2009})}\BibitemShut {NoStop}%
\bibitem [{\citenamefont {Yonezawa}\ \emph {et~al.}(2013)\citenamefont
  {Yonezawa}, \citenamefont {Kajikawa},\ and\ \citenamefont
  {Maeno}}]{PhysRevLett.110.077003}%
  \BibitemOpen
  \bibfield  {author} {\bibinfo {author} {\bibfnamefont {S.}~\bibnamefont
  {Yonezawa}}, \bibinfo {author} {\bibfnamefont {T.}~\bibnamefont {Kajikawa}},
  \ and\ \bibinfo {author} {\bibfnamefont {Y.}~\bibnamefont {Maeno}},\ }\href
  {\doibase 10.1103/PhysRevLett.110.077003} {\bibfield  {journal} {\bibinfo
  {journal} {Phys. Rev. Lett.}\ }\textbf {\bibinfo {volume} {110}},\ \bibinfo
  {pages} {077003} (\bibinfo {year} {2013})}\BibitemShut {NoStop}%
\bibitem [{\citenamefont {Maeno}\ \emph {et~al.}(2012)\citenamefont {Maeno},
  \citenamefont {Kittaka}, \citenamefont {Nomura}, \citenamefont {Yonezawa},\
  and\ \citenamefont {Ishida}}]{JPSJ.81.011009}%
  \BibitemOpen
  \bibfield  {author} {\bibinfo {author} {\bibfnamefont {Y.}~\bibnamefont
  {Maeno}}, \bibinfo {author} {\bibfnamefont {S.}~\bibnamefont {Kittaka}},
  \bibinfo {author} {\bibfnamefont {T.}~\bibnamefont {Nomura}}, \bibinfo
  {author} {\bibfnamefont {S.}~\bibnamefont {Yonezawa}}, \ and\ \bibinfo
  {author} {\bibfnamefont {K.}~\bibnamefont {Ishida}},\ }\href {\doibase
  10.1143/JPSJ.81.011009} {\bibfield  {journal} {\bibinfo  {journal} {Journal
  of the Physical Society of Japan}\ }\textbf {\bibinfo {volume} {81}},\
  \bibinfo {pages} {011009} (\bibinfo {year} {2012})}\BibitemShut {NoStop}%
\bibitem [{\citenamefont {Haverkort}\ \emph {et~al.}(2008)\citenamefont
  {Haverkort}, \citenamefont {Elfimov}, \citenamefont {Tjeng}, \citenamefont
  {Sawatzky},\ and\ \citenamefont {Damascelli}}]{PhysRevLett.101.026406}%
  \BibitemOpen
  \bibfield  {author} {\bibinfo {author} {\bibfnamefont {M.~W.}\ \bibnamefont
  {Haverkort}}, \bibinfo {author} {\bibfnamefont {I.~S.}\ \bibnamefont
  {Elfimov}}, \bibinfo {author} {\bibfnamefont {L.~H.}\ \bibnamefont {Tjeng}},
  \bibinfo {author} {\bibfnamefont {G.~A.}\ \bibnamefont {Sawatzky}}, \ and\
  \bibinfo {author} {\bibfnamefont {A.}~\bibnamefont {Damascelli}},\ }\href
  {\doibase 10.1103/PhysRevLett.101.026406} {\bibfield  {journal} {\bibinfo
  {journal} {Phys. Rev. Lett.}\ }\textbf {\bibinfo {volume} {101}},\ \bibinfo
  {pages} {026406} (\bibinfo {year} {2008})}\BibitemShut {NoStop}%
\bibitem [{\citenamefont {Ng}\ and\ \citenamefont
  {Sigrist}(2000)}]{EPL_SO_Sigrist}%
  \BibitemOpen
  \bibfield  {author} {\bibinfo {author} {\bibfnamefont {K.~K.}\ \bibnamefont
  {Ng}}\ and\ \bibinfo {author} {\bibfnamefont {M.}~\bibnamefont {Sigrist}},\
  }\href@noop {} {\bibfield  {journal} {\bibinfo  {journal} {Europhys. Lett.}\
  }\textbf {\bibinfo {volume} {49}},\ \bibinfo {pages} {473} (\bibinfo {year}
  {2000})}\BibitemShut {NoStop}%
\bibitem [{\citenamefont {Raghu}\ \emph {et~al.}(2010)\citenamefont {Raghu},
  \citenamefont {Kapitulnik},\ and\ \citenamefont {Kivelson}}]{raghu}%
  \BibitemOpen
  \bibfield  {author} {\bibinfo {author} {\bibfnamefont {S.}~\bibnamefont
  {Raghu}}, \bibinfo {author} {\bibfnamefont {A.}~\bibnamefont {Kapitulnik}}, \
  and\ \bibinfo {author} {\bibfnamefont {S.~A.}\ \bibnamefont {Kivelson}},\
  }\href {\doibase 10.1103/PhysRevLett.105.136401} {\bibfield  {journal}
  {\bibinfo  {journal} {Phys. Rev. Lett.}\ }\textbf {\bibinfo {volume} {105}},\
  \bibinfo {pages} {136401} (\bibinfo {year} {2010})}\BibitemShut {NoStop}%
\bibitem [{\citenamefont {Deisz}\ and\ \citenamefont
  {Kidd}(2011)}]{PhysRevLett.107.277003}%
  \BibitemOpen
  \bibfield  {author} {\bibinfo {author} {\bibfnamefont {J.~J.}\ \bibnamefont
  {Deisz}}\ and\ \bibinfo {author} {\bibfnamefont {T.~E.}\ \bibnamefont
  {Kidd}},\ }\href {\doibase 10.1103/PhysRevLett.107.277003} {\bibfield
  {journal} {\bibinfo  {journal} {Phys. Rev. Lett.}\ }\textbf {\bibinfo
  {volume} {107}},\ \bibinfo {pages} {277003} (\bibinfo {year}
  {2011})}\BibitemShut {NoStop}%
\bibitem [{\citenamefont {{Puetter}}\ and\ \citenamefont
  {{Kee}}(2012)}]{arXiv1101.4656}%
  \BibitemOpen
  \bibfield  {author} {\bibinfo {author} {\bibfnamefont {C.~M.}\ \bibnamefont
  {{Puetter}}}\ and\ \bibinfo {author} {\bibfnamefont {H.-Y.}\ \bibnamefont
  {{Kee}}},\ }\href {\doibase 10.1209/0295-5075/98/27010} {\bibfield  {journal}
  {\bibinfo  {journal} {EPL}\ }\textbf {\bibinfo {volume} {98}},\ \bibinfo
  {pages} {27010} (\bibinfo {year} {2012})}\BibitemShut {NoStop}%
\bibitem [{\citenamefont {Damascelli}\ \emph {et~al.}(2000)\citenamefont
  {Damascelli}, \citenamefont {Lu}, \citenamefont {Shen}, \citenamefont
  {Armitage}, \citenamefont {Ronning}, \citenamefont {Feng}, \citenamefont
  {Kim}, \citenamefont {Shen}, \citenamefont {Kimura}, \citenamefont {Tokura},
  \citenamefont {Mao},\ and\ \citenamefont {Maeno}}]{PhysRevLett.85.5194}%
  \BibitemOpen
  \bibfield  {author} {\bibinfo {author} {\bibfnamefont {A.}~\bibnamefont
  {Damascelli}}, \bibinfo {author} {\bibfnamefont {D.~H.}\ \bibnamefont {Lu}},
  \bibinfo {author} {\bibfnamefont {K.~M.}\ \bibnamefont {Shen}}, \bibinfo
  {author} {\bibfnamefont {N.~P.}\ \bibnamefont {Armitage}}, \bibinfo {author}
  {\bibfnamefont {F.}~\bibnamefont {Ronning}}, \bibinfo {author} {\bibfnamefont
  {D.~L.}\ \bibnamefont {Feng}}, \bibinfo {author} {\bibfnamefont
  {C.}~\bibnamefont {Kim}}, \bibinfo {author} {\bibfnamefont {Z.-X.}\
  \bibnamefont {Shen}}, \bibinfo {author} {\bibfnamefont {T.}~\bibnamefont
  {Kimura}}, \bibinfo {author} {\bibfnamefont {Y.}~\bibnamefont {Tokura}},
  \bibinfo {author} {\bibfnamefont {Z.~Q.}\ \bibnamefont {Mao}}, \ and\
  \bibinfo {author} {\bibfnamefont {Y.}~\bibnamefont {Maeno}},\ }\href
  {\doibase 10.1103/PhysRevLett.85.5194} {\bibfield  {journal} {\bibinfo
  {journal} {Phys. Rev. Lett.}\ }\textbf {\bibinfo {volume} {85}},\ \bibinfo
  {pages} {5194} (\bibinfo {year} {2000})}\BibitemShut {NoStop}%
\bibitem [{\citenamefont {Shen}\ \emph {et~al.}(2001)\citenamefont {Shen},
  \citenamefont {Damascelli}, \citenamefont {Lu}, \citenamefont {Armitage},
  \citenamefont {Ronning}, \citenamefont {Feng}, \citenamefont {Kim},
  \citenamefont {Shen}, \citenamefont {Singh}, \citenamefont {Mazin},
  \citenamefont {Nakatsuji}, \citenamefont {Mao}, \citenamefont {Maeno},
  \citenamefont {Kimura},\ and\ \citenamefont {Tokura}}]{PhysRevB.64.180502}%
  \BibitemOpen
  \bibfield  {author} {\bibinfo {author} {\bibfnamefont {K.~M.}\ \bibnamefont
  {Shen}}, \bibinfo {author} {\bibfnamefont {A.}~\bibnamefont {Damascelli}},
  \bibinfo {author} {\bibfnamefont {D.~H.}\ \bibnamefont {Lu}}, \bibinfo
  {author} {\bibfnamefont {N.~P.}\ \bibnamefont {Armitage}}, \bibinfo {author}
  {\bibfnamefont {F.}~\bibnamefont {Ronning}}, \bibinfo {author} {\bibfnamefont
  {D.~L.}\ \bibnamefont {Feng}}, \bibinfo {author} {\bibfnamefont
  {C.}~\bibnamefont {Kim}}, \bibinfo {author} {\bibfnamefont {Z.-X.}\
  \bibnamefont {Shen}}, \bibinfo {author} {\bibfnamefont {D.~J.}\ \bibnamefont
  {Singh}}, \bibinfo {author} {\bibfnamefont {I.~I.}\ \bibnamefont {Mazin}},
  \bibinfo {author} {\bibfnamefont {S.}~\bibnamefont {Nakatsuji}}, \bibinfo
  {author} {\bibfnamefont {Z.~Q.}\ \bibnamefont {Mao}}, \bibinfo {author}
  {\bibfnamefont {Y.}~\bibnamefont {Maeno}}, \bibinfo {author} {\bibfnamefont
  {T.}~\bibnamefont {Kimura}}, \ and\ \bibinfo {author} {\bibfnamefont
  {Y.}~\bibnamefont {Tokura}},\ }\href {\doibase 10.1103/PhysRevB.64.180502}
  {\bibfield  {journal} {\bibinfo  {journal} {Phys. Rev. B}\ }\textbf {\bibinfo
  {volume} {64}},\ \bibinfo {pages} {180502} (\bibinfo {year}
  {2001})}\BibitemShut {NoStop}%
\bibitem [{\citenamefont {{Veenstra}}\ \emph {et~al.}(2013)\citenamefont
  {{Veenstra}}, \citenamefont {{Zhu}}, \citenamefont {{Ludbrook}},
  \citenamefont {{Capsoni}}, \citenamefont {{Levy}}, \citenamefont
  {{Nicolaou}}, \citenamefont {{Rosen}}, \citenamefont {{Comin}}, \citenamefont
  {{Kittaka}}, \citenamefont {{Maeno}}, \citenamefont {{Elfimov}},\ and\
  \citenamefont {{Damascelli}}}]{MineNew}%
  \BibitemOpen
  \bibfield  {author} {\bibinfo {author} {\bibfnamefont {C.~N.}\ \bibnamefont
  {{Veenstra}}}, \bibinfo {author} {\bibfnamefont {Z.-H.}\ \bibnamefont
  {{Zhu}}}, \bibinfo {author} {\bibfnamefont {B.}~\bibnamefont {{Ludbrook}}},
  \bibinfo {author} {\bibfnamefont {M.}~\bibnamefont {{Capsoni}}}, \bibinfo
  {author} {\bibfnamefont {G.}~\bibnamefont {{Levy}}}, \bibinfo {author}
  {\bibfnamefont {A.}~\bibnamefont {{Nicolaou}}}, \bibinfo {author}
  {\bibfnamefont {J.~A.}\ \bibnamefont {{Rosen}}}, \bibinfo {author}
  {\bibfnamefont {R.}~\bibnamefont {{Comin}}}, \bibinfo {author} {\bibfnamefont
  {S.}~\bibnamefont {{Kittaka}}}, \bibinfo {author} {\bibfnamefont
  {Y.}~\bibnamefont {{Maeno}}}, \bibinfo {author} {\bibfnamefont {I.~S.}\
  \bibnamefont {{Elfimov}}}, \ and\ \bibinfo {author} {\bibfnamefont
  {A.}~\bibnamefont {{Damascelli}}},\ }\href@noop {} {\bibfield  {journal}
  {\bibinfo  {journal} {Phys. Rev. Lett.}\ }\textbf {\bibinfo {volume} {110}},\
  \bibinfo {pages} {097004} (\bibinfo {year} {2013})}\BibitemShut {NoStop}%
\bibitem [{\citenamefont {Iwasawa}\ \emph {et~al.}(2010)\citenamefont
  {Iwasawa}, \citenamefont {Yoshida}, \citenamefont {Hase}, \citenamefont
  {Koikegami}, \citenamefont {Hayashi}, \citenamefont {Jiang}, \citenamefont
  {Shimada}, \citenamefont {Namatame}, \citenamefont {Taniguchi},\ and\
  \citenamefont {Aiura}}]{PhysRevLett.105.226406}%
  \BibitemOpen
  \bibfield  {author} {\bibinfo {author} {\bibfnamefont {H.}~\bibnamefont
  {Iwasawa}}, \bibinfo {author} {\bibfnamefont {Y.}~\bibnamefont {Yoshida}},
  \bibinfo {author} {\bibfnamefont {I.}~\bibnamefont {Hase}}, \bibinfo {author}
  {\bibfnamefont {S.}~\bibnamefont {Koikegami}}, \bibinfo {author}
  {\bibfnamefont {H.}~\bibnamefont {Hayashi}}, \bibinfo {author} {\bibfnamefont
  {J.}~\bibnamefont {Jiang}}, \bibinfo {author} {\bibfnamefont
  {K.}~\bibnamefont {Shimada}}, \bibinfo {author} {\bibfnamefont
  {H.}~\bibnamefont {Namatame}}, \bibinfo {author} {\bibfnamefont
  {M.}~\bibnamefont {Taniguchi}}, \ and\ \bibinfo {author} {\bibfnamefont
  {Y.}~\bibnamefont {Aiura}},\ }\href {\doibase 10.1103/PhysRevLett.105.226406}
  {\bibfield  {journal} {\bibinfo  {journal} {Phys. Rev. Lett.}\ }\textbf
  {\bibinfo {volume} {105}},\ \bibinfo {pages} {226406} (\bibinfo {year}
  {2010})}\BibitemShut {NoStop}%
\bibitem [{\citenamefont {Osterwalder}(2006)}]{LectNotePhys.697.95}%
  \BibitemOpen
  \bibfield  {author} {\bibinfo {author} {\bibfnamefont {J.}~\bibnamefont
  {Osterwalder}},\ }\href@noop {} {\bibfield  {journal} {\bibinfo  {journal}
  {Lect. Notes Phys.}\ }\textbf {\bibinfo {volume} {697}},\ \bibinfo {pages}
  {95} (\bibinfo {year} {2006})}\BibitemShut {NoStop}%
\bibitem [{\citenamefont {Damascelli}(2004)}]{Damascelli:physica}%
  \BibitemOpen
  \bibfield  {author} {\bibinfo {author} {\bibfnamefont {A.}~\bibnamefont
  {Damascelli}},\ }\href@noop {} {\bibfield  {journal} {\bibinfo  {journal}
  {Physica Scripta}\ }\textbf {\bibinfo {volume} {T109}},\ \bibinfo {pages}
  {61} (\bibinfo {year} {2004})}\BibitemShut {NoStop}%
\bibitem [{sup()}]{supplementary}%
  \BibitemOpen
  \href@noop {} {\bibinfo  {journal} {See Supplemental Material for methods,
  calculation details, additional experimental data, and the complete set of
  $\ex{\vec{s_k} \cdot \vec{s_{-k}}}$ results for the full $k_z$ range}\
  }\BibitemShut {NoStop}%
\bibitem [{not()}]{note_ls}%
  \BibitemOpen
\bibfield  {journal} {  }\href@noop {} {\bibinfo  {journal} {{We also note that
  $\ex{\vec{l}\cdot\vec{s}}=0$ does not necessarily imply that SO coupling is
  not important; while the eigenstates may still be entangled, the individual
  $\vec{l}\cdot\vec{s}$ vector components might simply sum to zero, as is the
  case for the bulges of the $\gamma$ band near the zone diagonals
  \cite{supplementary}}}\ }\BibitemShut {NoStop}%
\bibitem [{\citenamefont {Pierce}\ and\ \citenamefont
  {Meier}(1976)}]{PhysRevB.13.5484}%
  \BibitemOpen
\bibfield  {journal} {  }\bibfield  {author} {\bibinfo {author} {\bibfnamefont
  {D.~T.}\ \bibnamefont {Pierce}}\ and\ \bibinfo {author} {\bibfnamefont
  {F.}~\bibnamefont {Meier}},\ }\href {\doibase 10.1103/PhysRevB.13.5484}
  {\bibfield  {journal} {\bibinfo  {journal} {Phys. Rev. B}\ }\textbf {\bibinfo
  {volume} {13}},\ \bibinfo {pages} {5484} (\bibinfo {year}
  {1976})}\BibitemShut {NoStop}%
\bibitem [{\citenamefont {Mizokawa}\ \emph {et~al.}(2001)\citenamefont
  {Mizokawa}, \citenamefont {Tjeng}, \citenamefont {Sawatzky}, \citenamefont
  {Ghiringhelli}, \citenamefont {Tjernberg}, \citenamefont {Brookes},
  \citenamefont {Fukazawa}, \citenamefont {Nakatsuji},\ and\ \citenamefont
  {Maeno}}]{PhysRevLett.87.077202}%
  \BibitemOpen
  \bibfield  {author} {\bibinfo {author} {\bibfnamefont {T.}~\bibnamefont
  {Mizokawa}}, \bibinfo {author} {\bibfnamefont {L.~H.}\ \bibnamefont {Tjeng}},
  \bibinfo {author} {\bibfnamefont {G.~A.}\ \bibnamefont {Sawatzky}}, \bibinfo
  {author} {\bibfnamefont {G.}~\bibnamefont {Ghiringhelli}}, \bibinfo {author}
  {\bibfnamefont {O.}~\bibnamefont {Tjernberg}}, \bibinfo {author}
  {\bibfnamefont {N.~B.}\ \bibnamefont {Brookes}}, \bibinfo {author}
  {\bibfnamefont {H.}~\bibnamefont {Fukazawa}}, \bibinfo {author}
  {\bibfnamefont {S.}~\bibnamefont {Nakatsuji}}, \ and\ \bibinfo {author}
  {\bibfnamefont {Y.}~\bibnamefont {Maeno}},\ }\href {\doibase
  10.1103/PhysRevLett.87.077202} {\bibfield  {journal} {\bibinfo  {journal}
  {Phys. Rev. Lett.}\ }\textbf {\bibinfo {volume} {87}},\ \bibinfo {pages}
  {077202} (\bibinfo {year} {2001})}\BibitemShut {NoStop}%
\bibitem [{\citenamefont {Cooper}(1962)}]{cooper}%
  \BibitemOpen
  \bibfield  {author} {\bibinfo {author} {\bibfnamefont {J.~W.}\ \bibnamefont
  {Cooper}},\ }\href@noop {} {\bibfield  {journal} {\bibinfo  {journal} {Phys.
  Rev.}\ }\textbf {\bibinfo {volume} {128}},\ \bibinfo {pages} {681} (\bibinfo
  {year} {1962})}\BibitemShut {NoStop}%
\bibitem [{kra()}]{kramer}%
  \BibitemOpen
  \href@noop {} {\bibinfo  {journal} {Note that the two states of each
  Kramers-degenerate pair have opposite spin polarization, consistent with the
  absence of net spin polarization at any k point in
  ${\mathrm{Sr}}_{2}{\mathrm{RuO}}_{4}$}\ }\BibitemShut {NoStop}%
\bibitem [{\citenamefont {Hasan}\ and\ \citenamefont {Kane}(2010)}]{TI_RMP}%
  \BibitemOpen
\bibfield  {journal} {  }\bibfield  {author} {\bibinfo {author} {\bibfnamefont
  {M.}~\bibnamefont {Hasan}}\ and\ \bibinfo {author} {\bibfnamefont
  {C.}~\bibnamefont {Kane}},\ }\href {\doibase 10.1103/RevModPhys.82.3045}
  {\bibfield  {journal} {\bibinfo  {journal} {Rev. Mod. Phys.}\ }\textbf
  {\bibinfo {volume} {82}},\ \bibinfo {pages} {3045} (\bibinfo {year}
  {2010})}\BibitemShut {NoStop}%
\bibitem [{\citenamefont {Zhu}\ \emph {et~al.}(2013{\natexlab{a}})\citenamefont
  {Zhu}, \citenamefont {Veenstra}, \citenamefont {Levy}, \citenamefont
  {Ubaldini}, \citenamefont {Syers}, \citenamefont {Butch}, \citenamefont
  {Paglione}, \citenamefont {Haverkort}, \citenamefont {Elfimov},\ and\
  \citenamefont {Damascelli}}]{jason}%
  \BibitemOpen
  \bibfield  {author} {\bibinfo {author} {\bibfnamefont {Z.-H.}\ \bibnamefont
  {Zhu}}, \bibinfo {author} {\bibfnamefont {C.~N.}\ \bibnamefont {Veenstra}},
  \bibinfo {author} {\bibfnamefont {G.}~\bibnamefont {Levy}}, \bibinfo {author}
  {\bibfnamefont {A.}~\bibnamefont {Ubaldini}}, \bibinfo {author}
  {\bibfnamefont {P.}~\bibnamefont {Syers}}, \bibinfo {author} {\bibfnamefont
  {N.~P.}\ \bibnamefont {Butch}}, \bibinfo {author} {\bibfnamefont
  {J.}~\bibnamefont {Paglione}}, \bibinfo {author} {\bibfnamefont {M.~W.}\
  \bibnamefont {Haverkort}}, \bibinfo {author} {\bibfnamefont {I.~S.}\
  \bibnamefont {Elfimov}}, \ and\ \bibinfo {author} {\bibfnamefont
  {A.}~\bibnamefont {Damascelli}},\ }\href {\doibase
  10.1103/PhysRevLett.110.216401} {\bibfield  {journal} {\bibinfo  {journal}
  {Phys. Rev. Lett.}\ }\textbf {\bibinfo {volume} {110}},\ \bibinfo {pages}
  {216401} (\bibinfo {year} {2013}{\natexlab{a}})}\BibitemShut {NoStop}%
\bibitem [{\citenamefont {Zhu}\ \emph {et~al.}(2013{\natexlab{b}})\citenamefont
  {Zhu}, \citenamefont {Veenstra}, \citenamefont {Zhdanovich}, \citenamefont
  {Schneider}, \citenamefont {Okuda}, \citenamefont {Miyamoto}, \citenamefont
  {Zhu}, \citenamefont {Namatame}, \citenamefont {Taniguchi}, \citenamefont
  {Haverkort}, \citenamefont {Elfimov},\ and\ \citenamefont
  {Damascelli}}]{jasonnew}%
  \BibitemOpen
  \bibfield  {author} {\bibinfo {author} {\bibfnamefont {Z.-H.}\ \bibnamefont
  {Zhu}}, \bibinfo {author} {\bibfnamefont {C.~N.}\ \bibnamefont {Veenstra}},
  \bibinfo {author} {\bibfnamefont {S.}~\bibnamefont {Zhdanovich}}, \bibinfo
  {author} {\bibfnamefont {M.}~\bibnamefont {Schneider}}, \bibinfo {author}
  {\bibfnamefont {T.}~\bibnamefont {Okuda}}, \bibinfo {author} {\bibfnamefont
  {K.}~\bibnamefont {Miyamoto}}, \bibinfo {author} {\bibfnamefont {S.-Y.}\
  \bibnamefont {Zhu}}, \bibinfo {author} {\bibfnamefont {H.}~\bibnamefont
  {Namatame}}, \bibinfo {author} {\bibfnamefont {M.}~\bibnamefont {Taniguchi}},
  \bibinfo {author} {\bibfnamefont {M.~W.}\ \bibnamefont {Haverkort}}, \bibinfo
  {author} {\bibfnamefont {I.~S.}\ \bibnamefont {Elfimov}}, \ and\ \bibinfo
  {author} {\bibfnamefont {A.}~\bibnamefont {Damascelli}},\ }\href@noop {}
  {\bibfield  {journal} {\bibinfo  {journal} {submitted}\ } (\bibinfo {year}
  {2013}{\natexlab{b}})}\BibitemShut {NoStop}%
\bibitem [{\citenamefont {Bardeen}\ \emph {et~al.}(1957)\citenamefont
  {Bardeen}, \citenamefont {Cooper},\ and\ \citenamefont
  {Schrieffer}}]{PhysRev.108.1175}%
  \BibitemOpen
  \bibfield  {author} {\bibinfo {author} {\bibfnamefont {J.}~\bibnamefont
  {Bardeen}}, \bibinfo {author} {\bibfnamefont {L.~N.}\ \bibnamefont {Cooper}},
  \ and\ \bibinfo {author} {\bibfnamefont {J.~R.}\ \bibnamefont {Schrieffer}},\
  }\href {\doibase 10.1103/PhysRev.108.1175} {\bibfield  {journal} {\bibinfo
  {journal} {Phys. Rev.}\ }\textbf {\bibinfo {volume} {108}},\ \bibinfo {pages}
  {1175} (\bibinfo {year} {1957})}\BibitemShut {NoStop}%
\bibitem [{\citenamefont {Mukuda}\ \emph {et~al.}(2010)\citenamefont {Mukuda},
  \citenamefont {Ohara}, \citenamefont {Yashima}, \citenamefont {Kitaoka},
  \citenamefont {Settai}, \citenamefont {\ifmmode~\bar{O}\else \={O}\fi{}nuki},
  \citenamefont {Itoh},\ and\ \citenamefont {Haller}}]{PhysRevLett.104.017002}%
  \BibitemOpen
  \bibfield  {author} {\bibinfo {author} {\bibfnamefont {H.}~\bibnamefont
  {Mukuda}}, \bibinfo {author} {\bibfnamefont {T.}~\bibnamefont {Ohara}},
  \bibinfo {author} {\bibfnamefont {M.}~\bibnamefont {Yashima}}, \bibinfo
  {author} {\bibfnamefont {Y.}~\bibnamefont {Kitaoka}}, \bibinfo {author}
  {\bibfnamefont {R.}~\bibnamefont {Settai}}, \bibinfo {author} {\bibfnamefont
  {Y.}~\bibnamefont {\ifmmode~\bar{O}\else \={O}\fi{}nuki}}, \bibinfo {author}
  {\bibfnamefont {K.~M.}\ \bibnamefont {Itoh}}, \ and\ \bibinfo {author}
  {\bibfnamefont {E.~E.}\ \bibnamefont {Haller}},\ }\href {\doibase
  10.1103/PhysRevLett.104.017002} {\bibfield  {journal} {\bibinfo  {journal}
  {Phys. Rev. Lett.}\ }\textbf {\bibinfo {volume} {104}},\ \bibinfo {pages}
  {017002} (\bibinfo {year} {2010})}\BibitemShut {NoStop}%
\bibitem [{hea()}]{heavy}%
  \BibitemOpen
  \href@noop {} {\bibinfo  {journal} {In heavy-fermion compounds the SO
  coupling is much larger than the bandwidth, suppressing the momentum
  dependence of the pseudo-spin transform and enabling a pseudo-spin
  description of superconductivity}\ }\BibitemShut {NoStop}%
\end{thebibliography}%

\end{document}